\documentclass[onecolumn,secnumarabic,amssymb, nobibnotes, aps, prd]{revtex4-2}
\usepackage{amssymb,amsmath}
\usepackage{enumitem,kantlipsum}
\usepackage{lipsum}
\usepackage{soul}
\usepackage{cancel}
\usepackage{dcolumn}
\usepackage{gensymb}
\usepackage{graphicx}
\usepackage{psfrag}
\usepackage{color}
\usepackage{bm}
\usepackage[normalem]{ulem}
\usepackage{multirow}

\def\beq{\begin{equation}}
\def\eeq{\end{equation}}
\def\bea{\begin{eqnarray}}
\def\eea{\end{eqnarray}}
\usepackage{cleveref}
\usepackage{hyperref}
\hypersetup{
  colorlinks=true,
  citecolor=blue,
  linkcolor=cyan,
  urlcolor=blue,
}
\newcommand{\BibitemShut}[1]{}

\begin{document}
\title{Thermal effects in an imbalanced dipolar fermionic superfluid}
 \author{Subhanka Mal$^1$, Hiranmaya Mishra$^2$, Prasanta K. Panigrahi$^3$, and Bimalendu Deb$^1$}
 \affiliation{$^1$School of Physical Sciences, Indian Association for the Cultivation of Science, Jadavpur, Kolkata 700032, India.\\
 $^2$School of Physical Sciences, National Institute of Science Education and Research, Jatni, Khurda 752050, Odisha, India.\\
 $^3$Department of Physical Sciences, IISER Kolkata, Mohanpur, Nadia 741246, West Bengal, India.}
 
\begin{abstract}
  We investigate the temperature effects in an imbalanced superfluid atomic Fermi gas. We consider a bilayer system of two-component dipolar fermionic atoms with one layer containing atoms of one component and the other layer the atoms of other component with an imbalance between the populations of the two components. This imbalance results in uniform and nonuniform superfluid phases such as phase-separated BCS, Fulde-Ferrel-Larkin-Ovchinnikov (FFLO), Sarma and normal Fermi liquid phases for different system parameters. Using the mean-field BCS theory together with the superfluid mass-density criterion we classify different phases in thermodynamic phase diagram. Our results indicate that for a dipolar Fermi system the Sarma phase is stable for large imbalance at finite temperature below the critical temperature, and the FFLO phase is stable for intermediate imbalance on the BCS side of a BCS-BCE crossover. The phase diagram in the temperature and population imbalance plane indicate three Lifshitz points: one corresponding to coexistance of BCS, FFLO and normal Fermi liquid phase while the other two correspond to the coexistance of the Sarma phase, FFLO phase and normal Fermi phase for dipolar interactions. 
\end{abstract}

\maketitle

\vspace{0.2in}
\section{Introduction}
Realizing and understanding various facets of fermionic superfluidity (FS) is an important pursuit in condensed matter physics. Superfluidity of charged fermions is what is known as superconductivity which is closely related to the superfluidity of neutral Fermi systems. Low temperature FS and superconductivity are generally described by Bardeen-Cooper-Schrieffer (BCS) theory \cite{BCS_PR:1957}. Depending on the range of coupling strengths, interactions, dimensionality, and other physical conditions, a Fermi system may exhibit a host of superfluid phenomena. Among the neutral Fermi systems, liquid $^3$He and ultracold Fermi gases of atoms are the important ones for exploring FS. While liquid $^3$He exhibits anisotropic $p$-wave superfluidity, atomic Fermi gases are a promising experimental platform to simulate FS of different kinds. BCS states of a neutral atomic Fermi gas have been experimentally demonstrated\cite{Schirotzek_PRL:2008}.

Standard BCS theory is applicable to a Fermi system with two spin components (such as electrons, $^6$Li atomic gas, etc.) having equal number density for both components. However, interesting or exotic superfluid behavior may arise when there is an imbalance between the densities or masses of the two components, resulting in a mismatch between the two Fermi surfaces associated with the two components. This leads to one or more gapless modes in the excitation spectrum. Such a system with mismatched Fermi surfaces was first theoretically treated by Sarma \cite{Sarma_JPCS:1963} about 50 years ago. These phases are characterized as Sarma or breached-pair superfluid \cite{Forbes_PRL:2005} or interior gap phase \cite{Deb_PRA:2004}. Another nonuniform phase with mismatch between Fermi surfaces of the two components is known as Fulde-Ferrell-Larkin-Ovchinnikov phase \cite{Fulde_PR:1964,Larkin_ZETF:1965} where the Cooper pairs carry finite momentum and the superfluid gap has a spatially oscillating nature. In recent times, some of experiments using atomic Fermi gases with mismatched Fermi surfaces have been performed \cite{Ravensbergen_PRL:2020}. Though these experiments have revealed several new aspects of imbalanced Fermi systems, interior gap phase still remains elusive. Most of the experiments in cold atoms use tunable contact interactions over a wide range to understand the effects of coupling strength $k_Fa_s$ (where $k_F$ is the Fermi wave number and $a_s$ is the $s$-wave scattering length) on the FS.

The unprecedented control over various system parameters has indeed made ultracold atomic gases a unique platform to simulate numerous condensed matter phenomena, including FS. Recent experiments \cite{Jin_PRL:2004,Chin_RMP:2010} with strongly interacting Fermi gases have led to the observation of a crossover from BCS state of fermions to a Bose Einstein condensate (BEC) of bosonic molecules. These developments make Fermi gases a prospective system to explore exotic or unconvensional FS such as $p$-wave superfluidity \cite{Regal_PRL:2003, Zhang_PRA:2004, Ohashi_PRL:2005, Ho_PRL:2005, Gurarie_PRL:2005, Iskin_PRB:2005, Levinsen_PRL:2007, Inotani_PRA:2012}, FFLO \cite{Fulde_PR:1964, Larkin_ZETF:1965} and Sharma phase \cite{Sarma_JPCS:1963}, condensate of dipolar Fermi gases \cite{Baranov_PRA:2002, Bruun_PRL:2008, Ni_Nat:2010, Wu_PRL:2010, Heo_PRA:2012}, intermediate-temperature superfluidity \cite{Chien_PRL:2006}, population and mass-imbalanced Fermi superfluids \cite{Liu_PRL:2003, Wu_PRA:2003, Deb_PRA:2004, Lin_PRA:2006, Pao_PRA:2007, Parish_PRL:2007, Gubbels_PRL:2009, Baarsma_PRA:2010, Wang_SciRep:2017}. Over the last two decades experimentalists have been successful in realizing the population imbalance between two components\cite{Zwierlein_Sci:2006, Partridge_Sci:2006, Shin_Nat:2008} or two species \cite{Taglieber_PRL:2008, Wille_PRL:2008, Voigt_PRL:2009} of atomic Fermi gases. Although the exact nature of the superfluid state of imbalanced or polarized Fermi gas is still a subject of considerable debate, recent theoretical and experimental works have explored different phases of polarized Fermi gases in the mean-field limit at zero- \cite{Liu_PRL:2003, Wu_PRA:2003,Bedaque_PRL:2003} and finite temperatures \cite{Parish_PRL:2007,Parish_NatPhy:2007,Gubbels_PRL:2009,Wang_SciRep:2017}. From a theoretical point of view, the imbalanced Fermi systems with contact interaction can exhibit a rich phase diagram and phase transition between superfluid and normal states through a Lifshitz point \cite{Parish_NatPhy:2007,Midtgaard_PRA:2018,Zdybel_PRA:2021}. Finite temperature model of a two-dimensional imbalanced Fermi system with contact \cite{Toniolo_PRA:2017} interaction has been studied.

In recent times, imbalanced Fermi gases have been theoretically investigated in reduced dimensions with long range interactions. A BCS-BEC crossover in a bilayer dipolar Fermi gas has been predicted \cite{Zinner_PRA:2012}. The Sarma phase has been predicted to be stable on the BEC side of a BCS-BEC crossover \cite{Gubankova_PRB:2006,Pao_PRB:2006}. Mass-imbalanced Fermi mixtures of alkali and dipolar atomic gases \cite{Ravensbergen_PRA:2018, Ravensbergen_PRL:2020,Neri_PRA:2020} have been experimentally realized. Spin imbalanced systems in presence of long-range boson-mediated interaction \cite{Midtgaard_PRA:2018} and dipolar interaction \cite{Mazloom_PRB:2017} in a bilayer geometry under mean-field approximation have also been studied at zero temperature. The effects of anisotropy are distinctly reflected as the superfluid order parameter becomes momentum-dependent. To the best of our knowledge, all the studies with imbalanced bilayer dipolar Fermi gases have been studied at zero temperature. The thermal effects on bilayer imbalanced dipolar system have not been studied yet.

Here we investigate the thermodynamics of imbalanced Fermi gas in presence of dipolar interaction both in weak and strong coupling limits.  We adopt the mean-field approximation to study the system. We first discuss the finite temperature behavior of the standard BCS phase and non-uniform interior gap (IG) or Sarma phase near unitarity limit. Then we address a system of a two-component dipolar system in bilayer geometry. The bilayer structure is particularly favourable as it offers a proper stabilization to the Cooper pair formation with aniosotropic dipolar interaction. We compare the zero- and finite-temperature behavior of the dipolar system with the system with contact interaction. Our results show that in the presence of contact interaction the only stable phase is the standard BCS phase in the unitarity limit on the BCS side of a BCS-BEC crossover. The critical temperature as a function of interaction strength is shown for a fixed chemical potential imbalance. For imbalanced bilayer dipolar gas we show wave-number  dependence of the superfluid gap for different population imbalance and other system parameters. At finite temperatures in the BCS side of the crossover, we observe a stable FFLO and Sarma phase which are otherwise absent in case of contact interaction.

The outline of the paper is as follows: In Sec.\ref{sec.2} we present the formalism and describe the ground state of the system. In Sec.\ref{sec.2.1} we discuss the superfluid gap and density equations for the case of contact interaction within the mean field interaction at finite temperature. The mean-field model for a bilayer dipolar system is given in Sec.\ref{sec.2.2}. In section \ref{sec.3}, we illustrate our numerical results on the thermal effects in Fermi superfluidity. We conclude in Sec.\ref{sec.4}.

\section{The model and formalism} \label{sec.2}
The Hamiltonian for a two-species Fermi system is $H = H_0 + H_1$ where
\begin{eqnarray}
H_0 = \sum_{j=1,2}\int d{\bf r} \hat\Psi^\dagger_j({\bf r})\left[-\frac{\hbar^2}{2m_j}\nabla^2 - \mu_j({\bf r})\right]\hat\Psi_j({\bf r})
\label{eq.1}
\end{eqnarray} 
and
\begin{eqnarray}
H_1 = \frac{1}{2}\int \int d{\bf r}d{\bf r'}\hat\Psi^\dagger_1({\bf r})\hat\Psi^\dagger_2({\bf r'})V_{int}({\bf r},{\bf r'})\hat\Psi_2({\bf r'})\hat\Psi_1({\bf r})
\label{eq.2}
\end{eqnarray}
where $\mu_i$ the chemical potential of the $i$-th species and $V_{int}({\bf r},{\bf r'})$ the two-body interaction potential. $\hat\Psi_i({\bf r})$ represents the field annihilation operator of $i$-th species which can be written as 
\begin{eqnarray}
\hat\Psi_j({\bf r})=\frac{1}{\sqrt{\tau}}\sum_{\bf k}\exp(i{\bf k.r})\hat{c}_j({\bf k})
\label{eq.3}
\end{eqnarray}
where $\tau$ is the volume of the system. The two-body interaction $V_{int}({\bf r},{\bf r'})$ may be short range or long range or a combination of both. For a dilute system of weakly interacting atoms, one can safely disregard the effects of trap or apply the local density approximation (LDA). We now first develop the theory of imbalanced superfluidity at finite temperatures with a contact potential. Then we develop the theory of a bilayer dipolar system. 



\subsection{Imbalanced Fermi system with contact interaction} \label{sec.2.1}
The effective contact interaction can be expressed in terms of $s$-wave scattering length $a_s$ and the interaction potential in Eq.(\ref{eq.2}) can be written as $V_{int}({\bf r},{\bf r'})=g\delta({\bf r-r'})$ where $g=4\pi\hbar^2a_s/2\bar{m}$ is the two fermion coupling and $\bar{m}=m_1m_2/(m_1+m_2)$ is the reduced mass. One can write down the mean field Hamiltonian considering the variational ansatz for the fermionic pairing state \cite{Deb_PRA:2004}
\begin{eqnarray}
|\Omega\rangle = \exp\left[\frac{1}{2}\int[\hat{c}_j({\bf k})^\dagger f({\bf k})\hat{c}_k(-{\bf k})^\dagger]\epsilon_{jk}d{\bf k}-{\it H.c.}\right]|0\rangle
\label{eq.4}
\end{eqnarray}
where $|0\rangle$ is the vacuum state, $\hat c_j (\hat c_j^\dagger)$ is annihilation (creation) operator with $\hat c_j|0\rangle = 0$ and $f(k)$ is the condensation function. At zero temperature ($T=0$), the thermodynamic free energy is given by the 
\begin{eqnarray}
{\Omega} &=& \sum_j \langle\Omega|\mathcal{H}-\mu_j\psi^\dagger_j\psi_j|\Omega\rangle \nonumber\\
&=& \frac{1}{(2\pi)^3}\int d^3k \left[\bar\xi - \sqrt{\Delta^2+\bar\xi^2} +\omega_1 \Theta(-\omega_1) + \omega_2 \Theta(-\omega_2) \right] -\frac{\Delta^2 \bar{m}}{2\pi a_s}
\label{eq.5}
\end{eqnarray}
where $\mathcal{H}$ is the Hamiltonian density. Minimization of the thermodynamic potential with respect to $\Delta$ gives the mean field gap equation \cite{Deb_PRA:2004}
\begin{eqnarray}
1 = -\frac{2\pi a_s}{\bar m}\frac{1}{(2\pi)^3}\int d^3k \frac{1-\Theta(-\omega_1)-\Theta(-\omega_2)}{2\sqrt{\Delta^2+\bar\xi^2}}
\label{eq.6}
\end{eqnarray}
where $\epsilon_1=\hbar^2k^2/2m_i$ and $\omega_{1,2} = \sqrt{\bar\xi^2+\Delta^2} \pm \delta\epsilon \mp \delta\nu$ with $\delta\epsilon = (\epsilon_1-\epsilon_2)/2$ and $\delta\nu=(\nu_1-\nu_2)/2$. Here $\nu_i = \mu_i - gn_j|\epsilon_{ij}|$ is the effective chemical potential of the $i$th component, $\bar\xi = \bar\epsilon_1 - \bar\nu_1 $ ($\bar\epsilon=(\epsilon_1+\epsilon_2)/2$, $\bar\nu=(\nu_1+\nu_2)/2$) and $\Delta = -\frac{2\pi a_s}{\bar m}\sum_k\langle c_1({\bf -k}) c_2({\bf k})\rangle$ is the superfluid gap. Further, $\Theta(\omega_i)$ is the heaveside theta function denoting the distribution of $i$-th species at zero temperature. The gap equation (\ref{eq.6}) is obviously divergent and needs to be renormalized. This can be done by defining the renormalized coupling after subtracting the zero temperature and zero density contribution. Thus the gap equation becomes 
\begin{eqnarray}
1 = -\frac{2\pi a_s}{\bar m}\frac{1}{2\pi^3}\int d^3k \left[\frac{1-\Theta(-\omega_1)-\Theta(-\omega_2)}{2\sqrt{\Delta^2+\bar\xi^2}} - \frac{1}{2\bar\epsilon}\right]
\label{eq.7}
\end{eqnarray}
The density equations are calculated from Eq.(\ref{eq.5}) using $\rho_i=-\frac{\partial\Omega}{\partial\mu_i}$ as 
\begin{eqnarray}
\rho_{1(2)}=\frac{1}{(2\pi)^3}\int d^3k\left[\Theta(-\omega_{1(2)})+\frac{1}{2}\left(1-\frac{\bar\xi}{\sqrt{\Delta^2+\bar\xi^2}}\right)(1-\Theta(-\omega_{1(2)})-\Theta(-\omega_{2(1)}))\right]
\label{eq.8}
\end{eqnarray}
\begin{figure}[b]
\centering
\includegraphics[height=2.1in,width=4.6in]{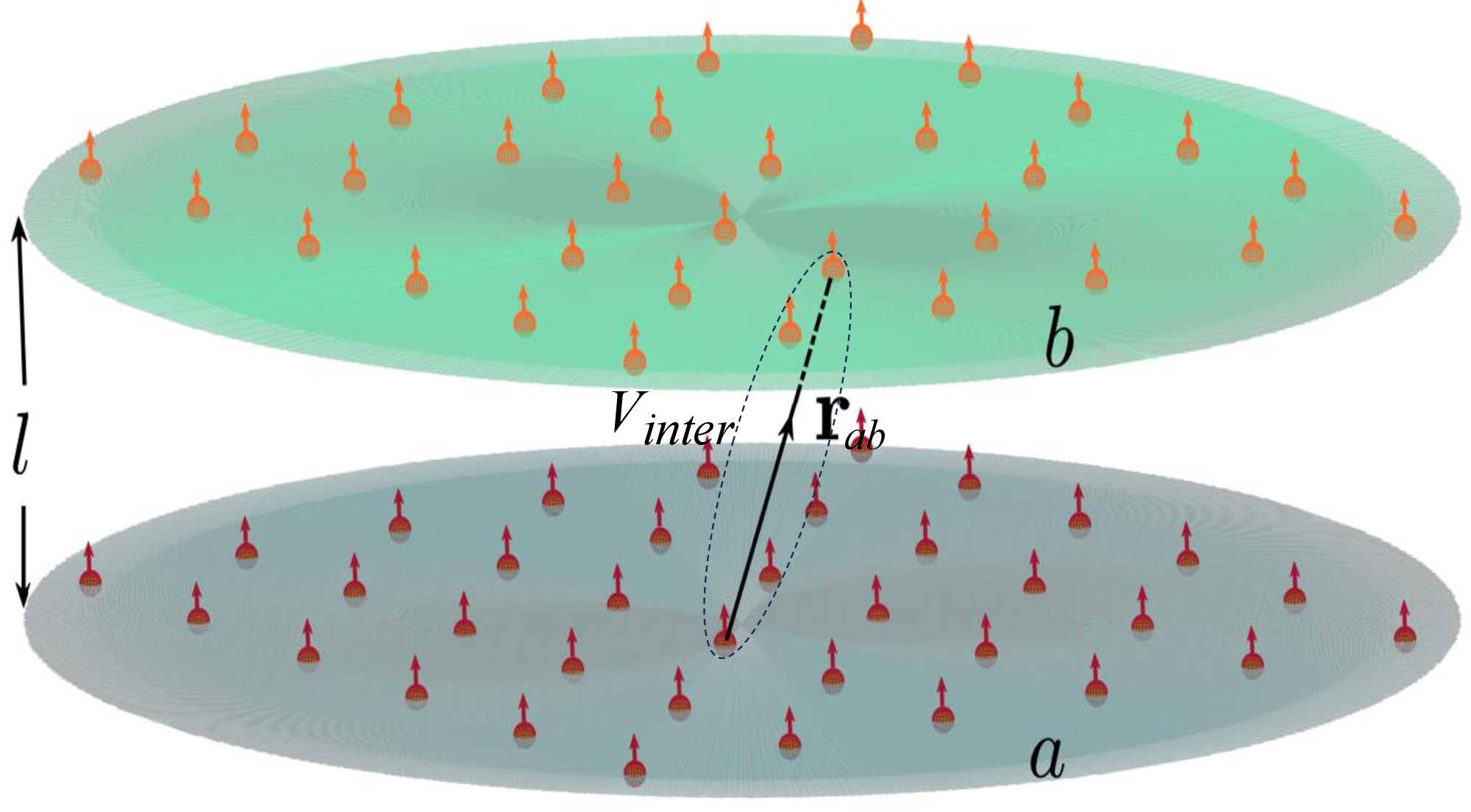}
 \caption{Pictorial representation of a two-dimensional bilayer imbalanced Fermi system. The interlayer spacing is $l$. The average dipole moment of each dipolar atom in both the layers is perpendicular to the surface of each layer.}
 \label{fig.1.2}
\end{figure}
For finite temperature the thermodynamic potential is
\begin{eqnarray}
{\Omega}(T) = \sum_j\left(\langle\Omega|\mathcal{H}-\mu_jN_j - \frac{1}{\beta}s_j|\Omega\rangle\right)
\label{eq.9}
\end{eqnarray}
where, $s_j=\frac{1}{(2\pi)^3}\sum_j\int d^3k[\theta_j\log(\theta_j) + (1-\theta_j)\log(1-\theta_j)]$ is the thermodynamic entropy density with $\theta_j(\omega_j)=(1+e^{\beta\omega_j})^{-1}$ being the thermal distribution function for the $j$-th species and $\beta = (k_BT)^{-1}$. At finite temperatures the thermodynamic potential takes the following form
\begin{eqnarray}
{\Omega}(T) = -\frac{\Delta^2\bar m}{2\pi a_s} +\frac{1}{(2\pi)^3}\int d^3k\left[\bar\xi+\sqrt{\Delta^2+\bar\xi^2}+ \frac{\Delta^2}{\bar\epsilon}\right]- \frac{1}{(2\pi)^3\beta} \int d^3k \Large[\log(1+e^{\beta\omega_1})\nonumber\\+\log(1+e^{\beta\omega_2})\Large]
\label{eq.10}
\end{eqnarray}
which essentially modifies the gap equation at finite temperature
\begin{eqnarray}
1 = -\frac{2\pi a_s}{\bar m}\frac{1}{(2\pi)^3}\int d^3k \left[\frac{1-\theta_1(\omega_1)-\theta_2(\omega_2)}{2\sqrt{\Delta^2+\bar\xi^2}} - \frac{1}{2\bar\epsilon}\right],
\label{eq.11}
\end{eqnarray}
and density equation
\begin{eqnarray}
\rho_{1(2)}=\frac{1}{(2\pi)^3}\int d^3k\left[\theta(-\omega_{1(2)})+\frac{1}{2}\left(1-\frac{\bar\xi}{\sqrt{\Delta^2+\bar\xi^2}}\right)\left(1-\theta(-\omega_{1(2)})-\theta(-\omega_{2(1)})\right)\right].
\end{eqnarray}

\subsection{Bilayer dipolar system} \label{sec.2.2}
In this section, we explore the properties of imbalanced superfluid in presence of long-range dipolar interaction. The effective interaction potential $V_{int}({\bf r, r'})$ can have one short range part and a long range part. In case of bosonic systems, both the parts are important but for the fermionic system only the long range part contributes to the low energy scattering \cite{Ronen_PRA:2006, Baranov_CR:2012}. In three dimensions the dipolar interaction is anisotropic and the resultant interaction strength depends on the angle between the dipole moment and the separation vector \cite{Baranov_CR:2012}. Depending on the angle the interaction can be either attaractive or repulsive. The attractive part of the dipolar interaction can induce instabilities in many-body systems \cite{zuchowski_PRA:2010}. Stabilization of dipolar interaction against inelastic collisions has been studied in Ref. \cite{Baranov_CR:2012}. It has been observered that a two-dimensional (or quasi-2D) {\it head-to-tail} configuration with an axial confinement leads to a stable dipolar Fermi system. For this purpose we consider a two-component bilayer dipolar Fermi system with one layer containing one type of fermionic atoms and the other layer with the other type. We also consider that there is a density imbalance between the two components. A schematic diagram of our model is shown in Fig.\ref{fig.1.2} with $l$ being the separation between the layers $a$ and $b$ and there is a population imbalance between the two layers. The effective Hamiltonian for such a system with the dipoles aligned perpendicular to the plane is given by
\begin{eqnarray}
H=\sum_{\bf k} \xi_k^a \hat{a}_{\bf k}^\dagger\hat{a}_{\bf k} + \sum_{\bf k} \xi_k^b \hat{b}_{\bf k}^\dagger\hat{b}_{\bf k} + \frac{1}{2}\sum_{\bf q} V_{\rm intra}(q) \left(\hat{a}_{\bf k+q}^\dagger\hat{a}_{\bf k-q}^\dagger\hat{a}_{\bf k}\hat{a}_{\bf k} + \hat{b}_{\bf k+q}^\dagger\hat{b}_{\bf k-q}^\dagger\hat{b}_{\bf k}\hat{b}_{\bf k} \right) \nonumber\\ + \sum_{\bf q} V_{\rm inter}(q) \hat{a}_{\bf k+q}^\dagger\hat{b}_{\bf k-q}^\dagger\hat{b}_{\bf k}\hat{a}_{\bf k}
\label{eq.12}
\end{eqnarray}
where $a_{\bf k}(a_{\bf k}^\dagger)$ and $b_{\bf k}(b_{\bf k}^\dagger)$ are annihilation (creation) operator in layer $a$ and $b$, respectively, $\xi_k^{a(b)} = \hbar^2k^2/(2m_{a(b)})-\mu_{a(b)}$, $V_{\rm intra}(q)$ and $V_{\rm inter}(q)$ are the intralayer and interlayer dipolar interaction in momentum space. Considering that the dipoles are oriented perpendicular to the plane, the intra- and interlayer dipole-dipole interactions in real space are $V_{\rm intra}(r) = \frac{C_{dd}}{4\pi}\frac{1}{r^3}$ and $V_{\rm inter}(r) = \frac{C_{dd}}{4\pi}\frac{r^2-2l^2}{(r^2+l^2)^{\frac{5}{2}}}$, respectively.
Here, $C_{dd}$ is the strength of dipole-dipole interaction, $l$ is the interlayer separation and $r$ is the in plane distance between two interacting dipoles. Let us note that the system has three length scales: the characteristic dipolar length $r_0 = \frac{C_{dd}m}{4\pi\hbar^2}$, the interlayer separation $l$ and the average interparticle separation in each layer $k_F^{-1} = 1/\sqrt{2\pi \rho}$ with $\rho = (\rho_a+\rho_b)$ being the total number of dipoles in the system. Now for a quasi-2D configuration, assuming only the ground state occupation is in the transverse direction, one obtains \cite{Zinner_PRA:2012,Baranov_CR:2012} the Fourier transform
\begin{eqnarray}
V_{\rm intra}(q) = \frac{C_{dd}}{4} \left[\frac{8}{3\sqrt{2\pi}w}-2qe^{q^2w^2/2}Erfc\left(\frac{qw}{\sqrt{2}}\right)\right]
\label{eq.13}
\end{eqnarray}
and
\begin{eqnarray}
V_{\rm inter}(q) = -\frac{C_{dd}}{2}qe^{-ql}
\label{eq.14}
\end{eqnarray}
in the limit where the width ($w$) of each layer is much less as compared to the interlayer separation ($l$). Here ${\bf q}$ is the in-plane momentum and ${Erfc}(x)=\frac{1}{2\pi}\int_x^\infty e^{-t^2}dt$ is the complementary error function.

In the BCS mean-field approximation considering only the $s$-wave pairing, the $k$-dependent superfluid order parameter is obtained from the following self-consistent equation
\begin{eqnarray}
    \Delta_k = -\frac{1}{2}\int V_{\rm inter}(|{\bf k - k'}|)\frac{\Delta_{k'}}{E_{k'}}{\big(}1 - \theta(\omega_a) - \theta(\omega_b){\big)} \frac{d^2k'}{(2\pi)^2}
    \label{eq.15}
\end{eqnarray}
where $E_k = \sqrt{\bar{\xi}_k^2+\Delta_k^2}$, $\bar{\xi}_k = \frac{1}{2}(\xi_k^{a} + \xi_k^{b})$ and $\omega_{a,b} = E_k \pm \frac{1}{2}(\xi_k^a - \xi_k^b)$. $\theta(\omega_i) = 1/[1+exp(\beta\omega_i)]$ is the Fermi-Dirac distribution function at inverse temperature $\beta = 1/(k_BT)$. Now, considering fixed density of dipole in each layer, the density equation for each layer can be represented as $\rho_{a(b)} = \int n_{a(b)}\frac{d^2k}{(2\pi)^2}$ where,
\begin{eqnarray}
n_{a(b)} = \frac{1}{2}{\Bigg[}\left(1+\frac{\bar\xi_k}{E_k}\right)\theta(\omega_{a(b)}) + \left(1-\frac{\bar\xi_k}{E_k}\right){\big(}1-\theta(\omega_{b(a)}){\big)}{\Bigg]}\nonumber
\end{eqnarray}




\section{Results and discussion} \label{sec.3}
For numerical illustration it is convenient to use dimensionless quantities in scale of average Fermi momentum $k_F$, Fermi energy $E_F$ and Fermi temperature $T_F$.   
As we discussed earlier, the superfluid gap can be obtained by minimizing the thermodynamic potential $\Omega$ with respect to the order parameter $\Delta$. Though our main focus of study is an imbalanced dipolar Fermi gas in a bilayer geometry, we first discuss the thermal effects in an imbalanced homogeneous Fermi gas with contact interaction for the sake of comparison between the two systems.
\begin{figure}[h]
\centering
\includegraphics[height=2.6in,width=6.4in]{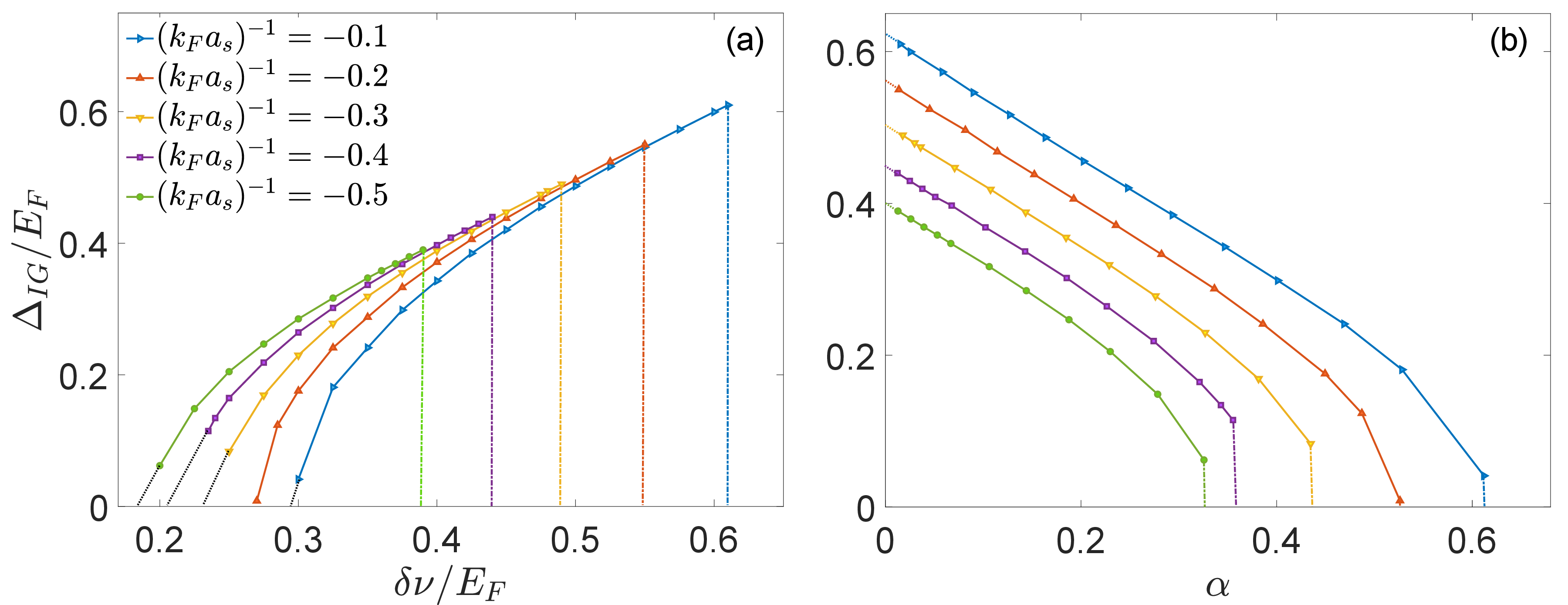}
 \caption{Variation of IG solution at zero temperature as a function of chemical potential imbalance $\delta\nu$ (a) and density imbalance $\alpha$ (b) for different values of $(k_Fa_s)^{-1}$. The dashed lines represent the critical imbalance values $\delta\nu_c$ and $\alpha_c$.}
 \label{fig.1.0}
\end{figure}
\begin{figure}[h]
\centering
\includegraphics[height=3.0in,width=6.4in]{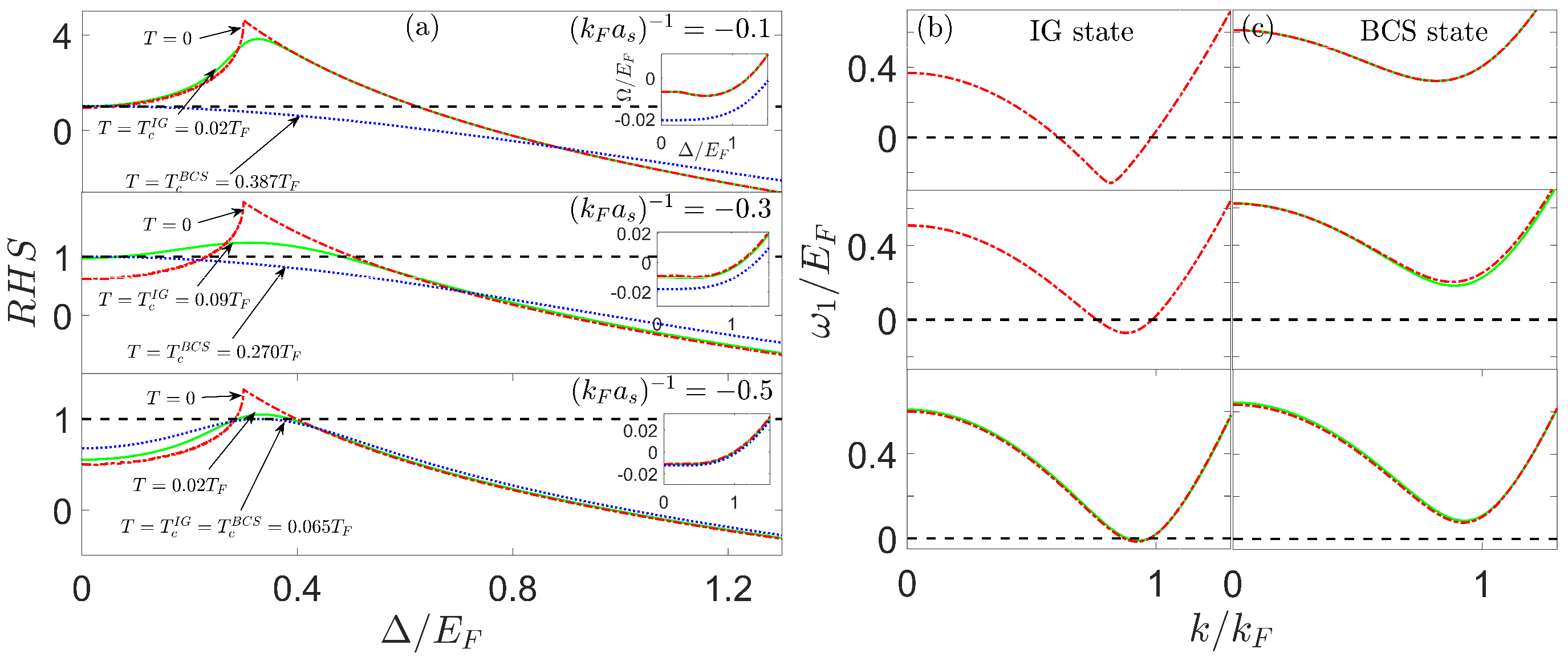}
 \caption{Right-hand side (RHS) of gap equation(s) and excitation spectrum are plotted as a function of dimensionless gap $\Delta/E_F$ and wave number $k/k_F$, respectively. (a) The variation of RHS with $\Delta/E_F$ for three different dimensionless temperatures $T/T_F$ and tree different $(k_Fa_s)^{-1}$. $\Delta$ is given by the condition RHS = 1. The smaller $\Delta$ corresponds to IG state and higher $\Delta$ corresponds to the BCS state. The insets show the dimensionless thermodynamic potential as a function of $\Delta/E_F$. The lower branch of excitation spectrum ($\omega_1$) in scale of $E_F$ for IG solution (b) and for BCS solution is shown as a function of $k/k_F$. Here $\delta\nu/E_F = 0.3$, $m_1=m_2=1$. }
 \label{fig.1.1}
\end{figure}
 \begin{figure}[htbp]
\centering
\includegraphics[height=2.5in,width=5.3in]{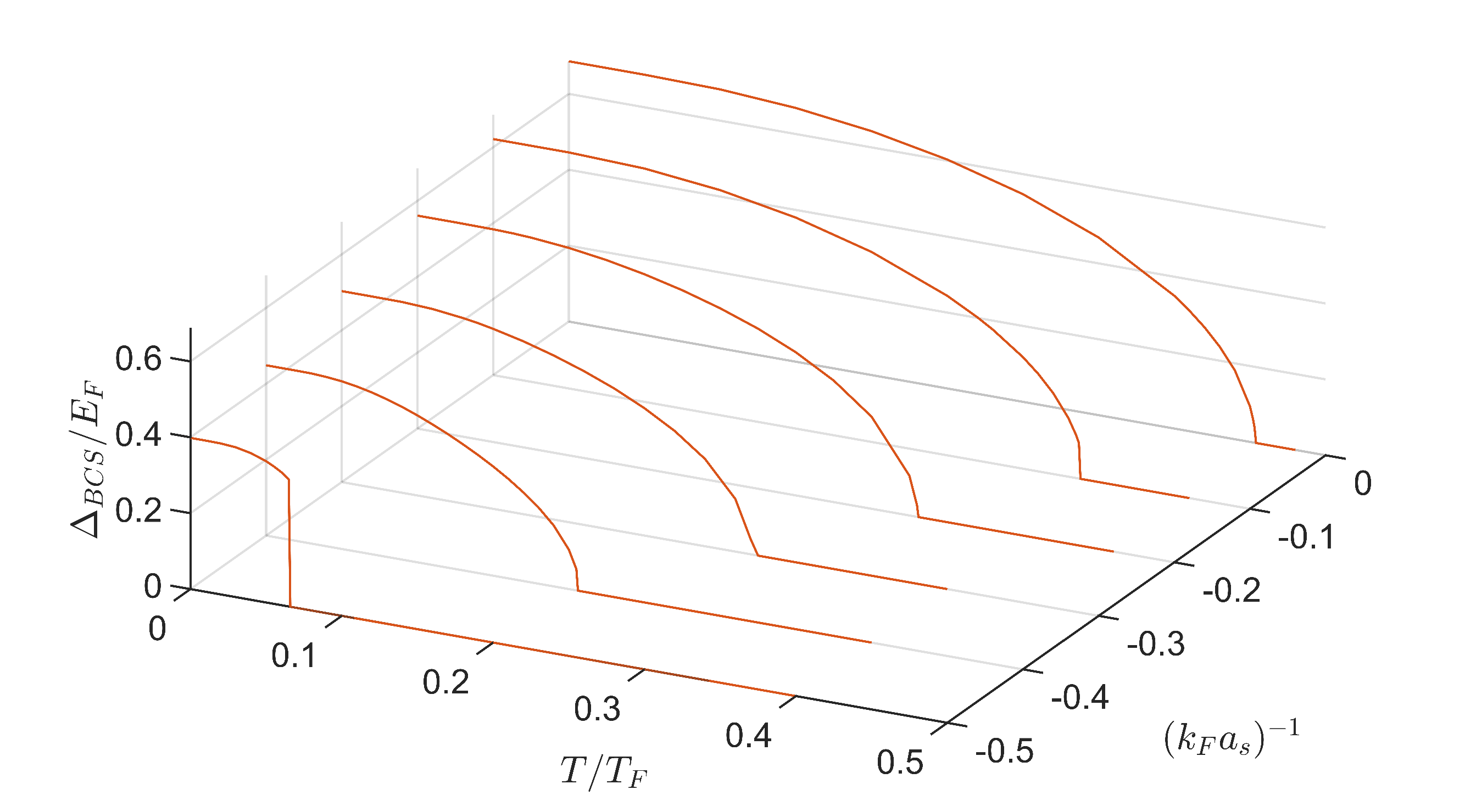}
 \caption{Variation of the dimensionless superfluid gap in BCS state $\Delta_{BCS}/E_F$ with 
dimensionless temperature $T/T_F$ for different interaction strength. Here $\delta\nu/E_F = 0.3$ and $m_1=m_2=1$. For each interaction strength which is quantified by the parameter $k_Fa_s$, there is a temperature after which the $\Delta_{BCS}$ vanishes, is the critical temperature at that interaction strength.}
 \label{fig.1.4}
\end{figure}

 \begin{figure}[htbp]
\centering
\includegraphics[height=2.5in,width=5.3in]{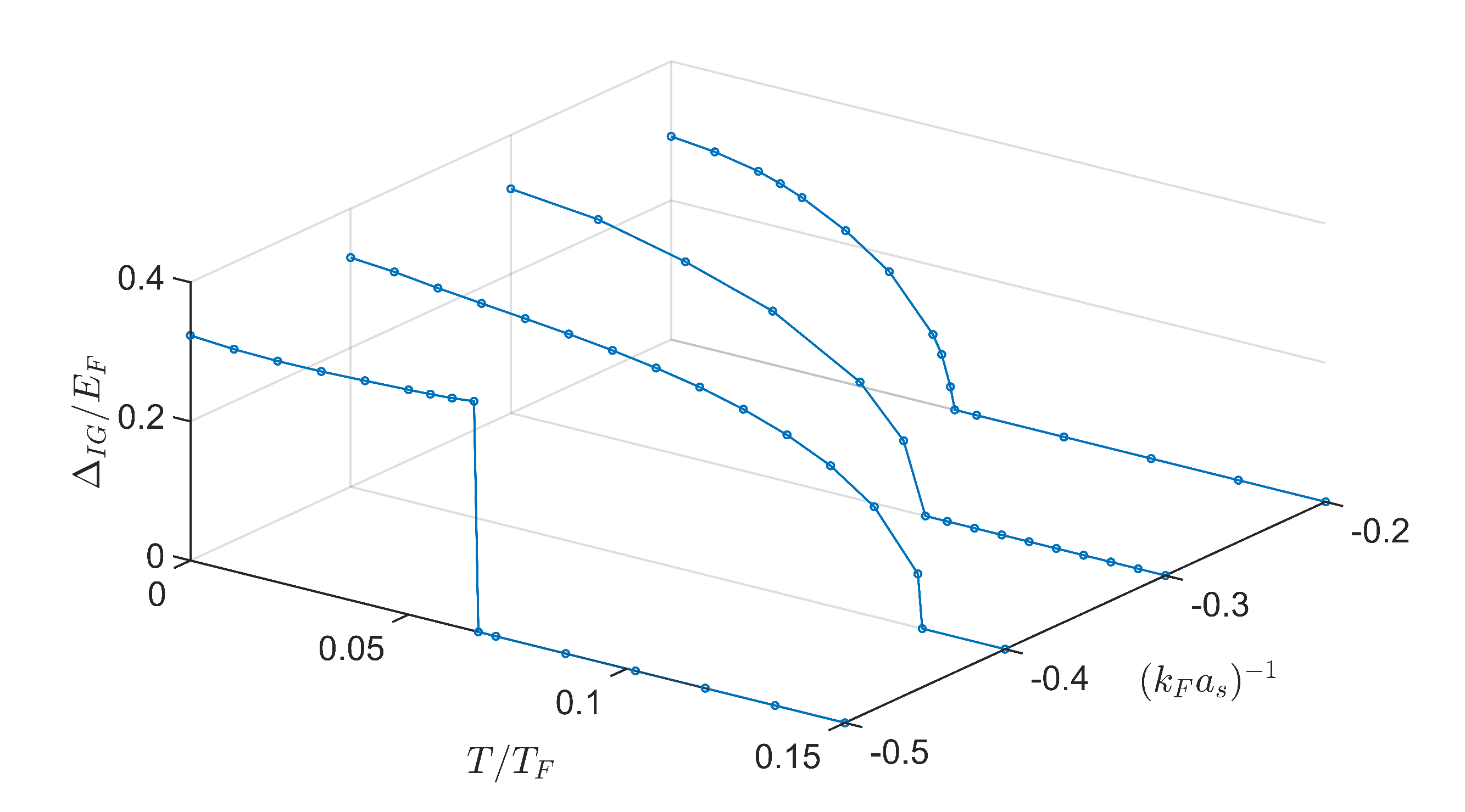}
 \caption{Variation of superfluid gap in IG state with temperature for different interaction strength keeping $\delta\nu/E_F = 0.3$.}
 \label{fig.1.5}
\end{figure}
\subsection{Imbalanced homogeneous Fermi gas with contact interaction}
 We first analyze the superfluid properties of a density imbalanced homogeneous Fermi gas with contact interaction. For our study, we consider the imbalance in chemical potential, which effectively provides the imbalance in density between two components. The masses of both the components are considered to be equal $m_1=m_2=m$. Considering a 3D homogeneous system the average Fermi momentum is $k_F=\left(3\pi^2(\rho_1+\rho_2)\right)^{\frac{1}{3}}$ and the corresponding average Fermi energy and Fermi temperature are $E_F=\hbar^2k_F^2/2m$ and $T_F=E_F/k_B$, respectively. The gap equation (\ref{eq.6}) at zero temperature is solved in consistent with the density equation (\ref{eq.8}). For the attractive interaction where $(k_Fa_s)^{-1}<0$, as the chemical potential imbalance $\delta\nu$ is increased, Eq.(\ref{eq.6}) has two solutions beyond a minimum value of $\delta\nu=\delta\nu_0$. Above $\delta\nu_0$, the solution with higher value for $\Delta$ is associated with the minimum of $\Omega$ in Eq.(\ref{eq.5}) with equal density $\rho_1=\rho_2$. This solution is characterized as the superfluid order parameter of the BCS state ($\Delta_{BCS}$). The lower solution for $\Delta$ corresponds to the maximum of $\Omega$ and is characterized as the interior gap (IG) phase or Sarma phase, and the associated solution is denoted as $\Delta_{IG}$. It is observed that for a particular value of interaction strength inverse $(k_Fa_s)^{-1}$ the BCS solution remains unchanged for all $\delta\nu>0$ up to a certain critical imbalance $\delta\nu_c=\Delta_{BCS}/\sqrt{2}$ which is associated with the Clogston-Chandrasekhar limit \cite{Clogston_PRL:1962}. However, as $\delta\nu$ crosses $\delta\nu_0$ $\Delta_{IG}$ becomes finite, and it increases with $\delta\nu$ till $\delta\nu_c$. We display the variation of $\Delta_{IG}$ as a function of $\delta\nu$ for different $(k_Fa_s)^{-1}$ in Fig.\ref{fig.1.0}(a) and with corresponding density imbalance $\alpha=(\rho_1-\rho_2)/(\rho_1+\rho_2)$ in Fig.\ref{fig.1.0}(b). $\delta\nu_c$ increases with decreasing $|k_Fa_s|^{-1}$. Consequently, the critical density imbalance $\alpha_c$ also increases as the interaction strength increases.
 
To discuss the thermal effects on the solutions of the superfluid gap equation, in Fig.\ref{fig.1.1}(a) we plot the RHS of Eq.(\ref{eq.11}) as a function of $\Delta/E_F$ for different values of the inverse of interaction strength $(k_Fa_s)^{-1}$ at zero as well as finite temperatures. The value of the chemical potential imbalance is $\delta\nu=0.3E_F$. It is observed from the figure that the gap $\Delta_{BCS}$ decreases and eventually vanishes at a critical temperature $T=T_c^{BCS}$. For a weaker coupling $(k_Fa_s)^{-1}=-0.5$, at $T=0$, $\Delta_{BCS}=0.399E_F$ and $\Delta_{IG}=0.304E_F$. As the temperature is increased, the BCS and IG solutions approach each other and vanish at $T=0.065T_F$. As the coupling is increased, e.g. $(k_Fa_s)^{-1}=-0.3$, the two gaps $\Delta_{IG}$ and $\Delta_{BCS}$ are $0.272E_F$ and $0.503E_F$ at zero temperature, respectively. With increasing temperature, the $\Delta_{IG}$ solution vanishes at $T=T_c^{IG}=0.09T_F$ and beyond that, only the $\Delta_{BCS}$ solution exists. As temperature is increased further, $\Delta_{BCS}$ vanishes at $T=T_c^{BCS}=0.27T_F$. For a stronger interaction, i.e. for $(k_Fa_s)^{-1}=-0.1$, $\Delta_{IG}$ becomes much smaller and $\Delta_{BCS}$ becomes larger. At $T=0$, $\Delta_{IG}=0.059E_F$ and $\Delta_{BCS}=0.622E_F$. As the temperature increases, $\Delta_{IG}$ vanishes at $T_c^{IG}=0.02T_F$ while $\Delta_{BCS}$ vanishes at $T_c^{BCS}=0.39T_F$.
As discussed earlier, the IG phase is unstable as it is associated with the maximum value of the thermodynamic potential as a function of $\Delta$ (shown in the inset of Fig.\ref{fig.1.1}(a)).

In Fig.\ref{fig.1.1}(b) and (c), we show the dispersion spectra of the quasi-particle energy $\omega_1$ as a function of $k/k_F$ for different temperatures and different interaction strengths. For the IG solution, there exist a pair of Fermi surfaces at $T=0$ where $\omega_1$ crosses zero. The corresponding Fermi momenta are $k_{F1}$ and $k_{F2}$. In between these points the the excitation remains gapless and the system is in the normal phase. For $(k_Fa_s)^{-1}=-0.1$, $k_{F1}=0.61k_F$ and $k_{F2}=0.98k_F$. As the strength of interaction decreases the gapless region shrinks signifying that the IG phase is absent below some certain critical interaction strength for a particular $\delta\nu$. On the other hand, for the BCS solution, there is no gapless mode in the excitation spectrum. In fact, as we show the nature of the RHS of Eq.(\ref{eq.6}), for a finite $\delta\nu>\delta\nu_0$, there is no solution for the superfluid gap below a critical interaction strength.

In Fig.\ref{fig.1.4} we show the evolution of BCS gap with temperature for different $(k_Fa_s)^{-1}$ with $\delta\nu = 0.3E_F$. It is observed that the critical temperature for the BCS state increases with increasing $|k_Fa_s|$. There are two different types of critical behavior. As discussed earlier, in the weakly interacting regime, we obtain a critical point for a finite $\Delta_{BCS}$, while in the strongly interacting regime, critical points are observed as $\Delta_{BCS}$ goes to zero.

The thermal behavior of the superfluid gap in the IG state is demonstrated in Fig.\ref{fig.1.5}. In contradiction to the BCS phase, the critical temperature in the IG phase decreases with increasing interaction strength in the strongly interacting limit. In the unitarity limit $(k_Fa_s)^{-1}\rightarrow0$, the IG solution ceases to exist even at zero temperature.

\subsection{Imbalanced bilayer dipolar system} \label{sec3.2}
Now we discuss the results for the imbalanced dipolar system. We consider a bilayer dipolar system with the alignment of the dipoles perpendicular to the plane as shown in Fig.\ref{fig.1.2}. The density imbalance $\alpha = (\rho_a-\rho_b)/(\rho_a+\rho_b)$ between two layers can be obtained by changing the chemical potential difference $h = (\mu_a-\mu_b)/2$. 
The existence non-zero solution of the gap equation is not necessarily associated with the uniform superfluid phase of the ground state. For $s$- wave pairing, the superfluid mass density \cite{Wu_PRA:2003,Iskin_PRL:2006,Pieri_PRB:2007,Subasi_PRB:2010}
\begin{eqnarray}
    \rho_s = m(\rho_a+\rho_b) - \frac{\hbar^2}{4\pi}\sum_j\frac{(k^a_j)^3}{|\frac{d\omega_a}{dk}|_{k^a_j}}
    \label{eq.17}
\end{eqnarray}
can also be used to characterise different superfluid phases in the ground state. Here $k^a_j$ is the $j$th zero of the quasiparticle dispersion $\omega_a(k)$. The superfluid mass density is composed of two parts \cite{Wu_PRA:2003}, one is paramagnetic and the other is diamagnetic. It is the competition between these two parts that determines the sign of the total mass density. A uniform superfluidity is characterized if there are no gapless mode in the excitation spectra and consequently $\rho_s=m(\rho_a+\rho_b)$. In presence of any gapless mode $\rho_s>0$ is a signature of the stable Sarma phase as it corresponds to a local minima of the thermodynamic potential. The negativity of $\rho_s$ can be characterised as the FFLO phase, where the effective current is in the opposite direction of the superfluid velocity due to the finite momentum of the Cooper pairs \cite{Mazloom_PRB:2017}.

\begin{figure}[htbp]
\centering
\includegraphics[height=2.1in,width=2.9in]{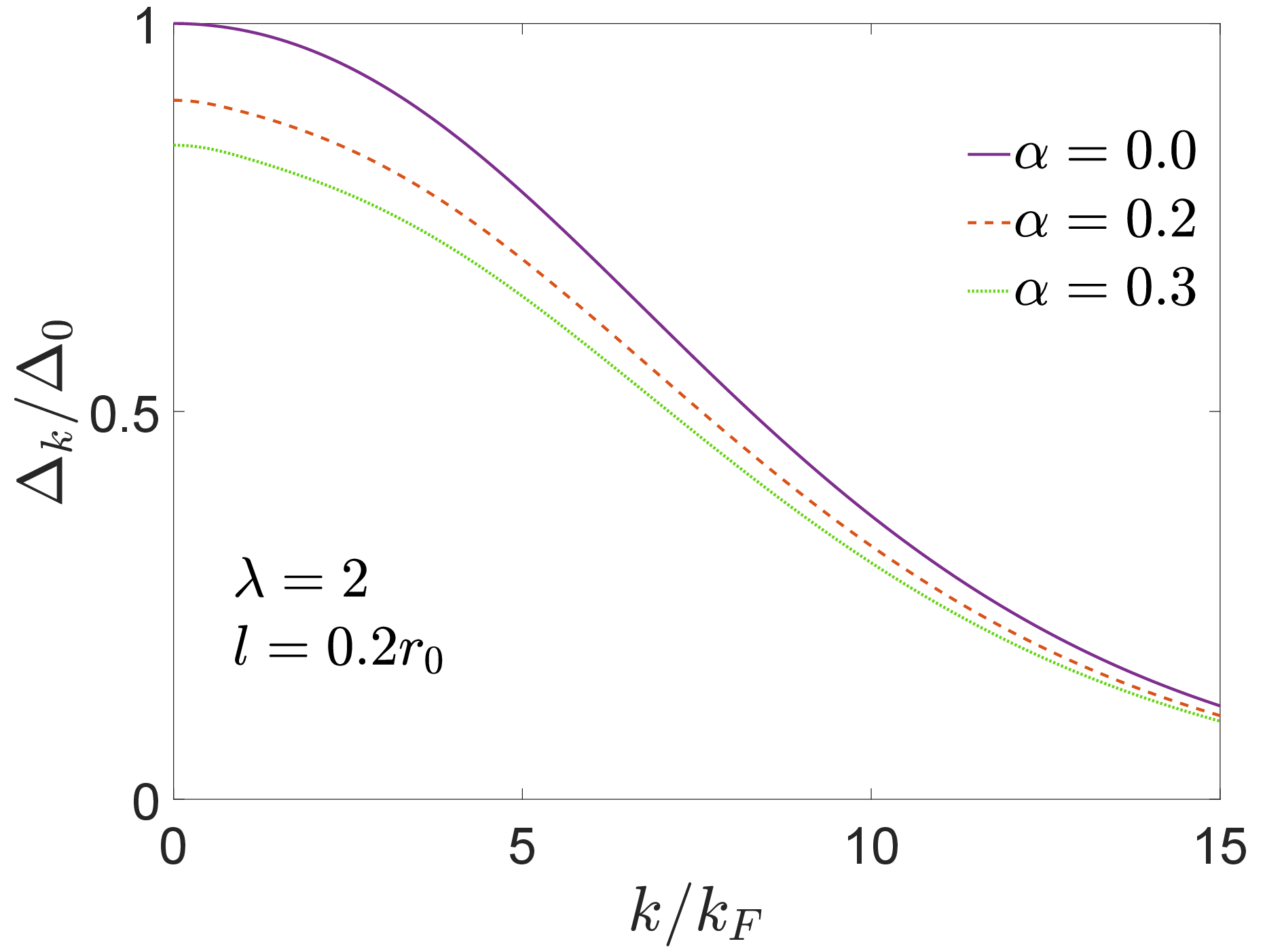}
 \caption{Comparison between $k$-dependent superfluid gaps of unpolarized bilayer systems with polarized ones. Here $\Delta_0$ is the pairing gap at $k = 0$ for unpolarized system.}
 \label{fig.1.6}
\end{figure}

 \begin{figure}[h !]
\centering
\includegraphics[height=2.1in,width=6.6in]{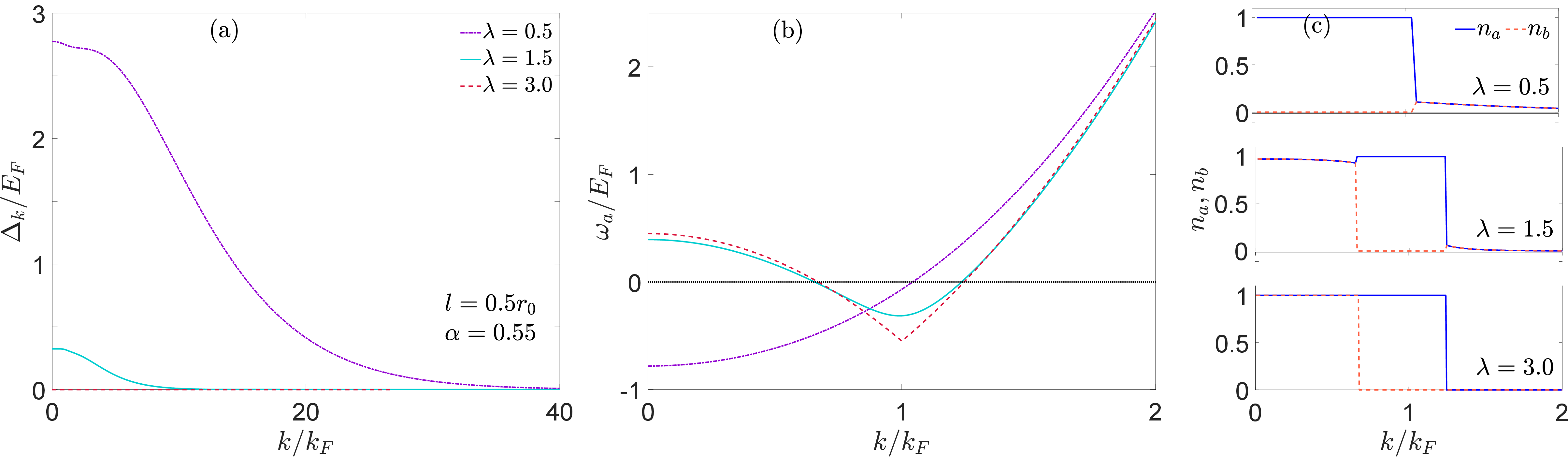}
 \caption{Zero-temperature $k$-dependence of superfluid gap $\Delta_k$ (a), lower branch of excitation spectrum $\omega_a$ (b) and density of fermions (c) are shown for a fixed density imbalance at an interlayer spacing $l = 0.5r_0$. For $\lambda = 0.5$, $1.5$ and $3$ we get Sarma, FFLO and normal Fermi liquid phases.}
 \label{fig.1.7}
\end{figure}

In case of bilayer system of dipolar superfluid, the pairing gap becomes $k$-(or energy) dependent. In Fig.\ref{fig.1.6} we show the $k$-dependence of the dimensionless superfluid gap $\Delta_k/\Delta_0$ for balanced and density imbalanced dipolar system at zero temperature for $\lambda=k_Fr_0=2$ and inter-layer separation $l=0.2r_0$, where $\Delta_0$ is the superfluid gap at $k=0$ and zero population imbalance. It is observed that as the imbalance increases, the amplitude of the gap $\Delta_k$ decreases. This is evident as the number of Cooper pairs reduces with increasing imbalance resulting in a phase-separated state of superfluid and normal Fermi liquid. It is worth mentioning that our results at zero temperature are in good agreement with those reported in Ref.\cite{Mazloom_PRB:2017}.

For a fixed density imbalance and inter-layer separation, we observe different phases with different total densities in Fig.\ref{fig.1.7} at zero temperature. For $\lambda =0.5$, the Sarma phase with single Fermi surface is observed. On the other hand, for $\lambda = 1.5$, there are two Fermi surfaces with negative mass density characteristic of a nonuniform FFLO phase. In  both the cases the interlayer separation $l=0.5r_0$ and density imbalance $\alpha=0.55$. As $\lambda$ increases, e.g. $\lambda = 3.0$ the system is in a normal Fermi liquid (NFL) phase which is also evident from Fig.\ref{fig.1.7}(a). We may note that the Sarma phase occurs in the BEC regime with a negative chemical potential $\mu=-2.35E_F$. Thus for $\lambda=0.5$, Sarma phase is stable only on the BEC side. On the other hand, as $\lambda$ increases, i.e., for $\lambda=1.5$ FFLO phase is stable in the BCS side with $\mu=0.98E_F$ which is consistent with Ref.\cite{Mazloom_PRB:2017}. Further increase in $\lambda$ destroys the superfluidity and the system becomes NFL with $\mu=E_F$.

\begin{figure}[htbp]
\centering
\includegraphics[height=2.5in,width=5.3in]{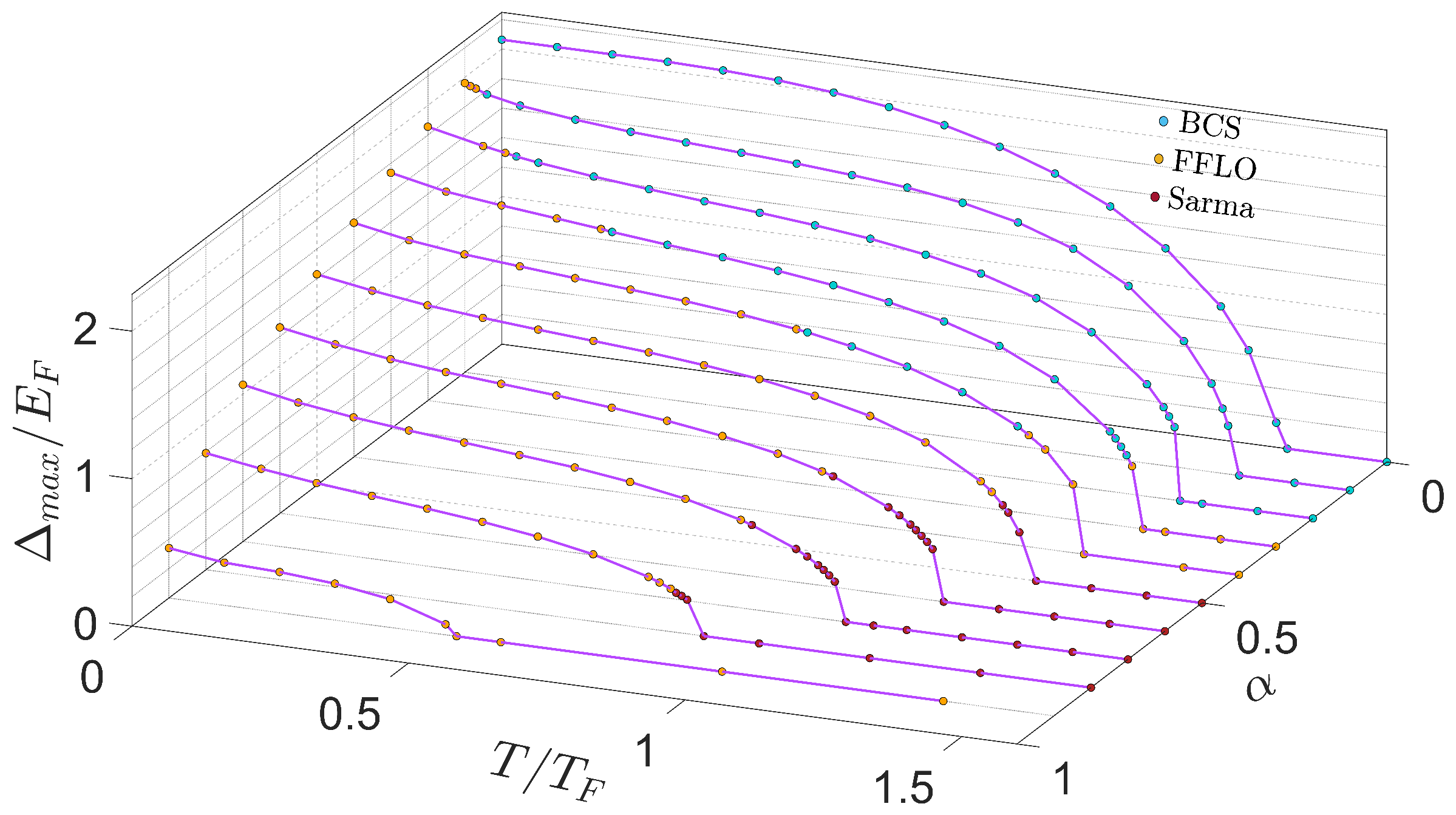}
 \caption{Finite temperature behavior of the dimensionless maximum superfluid gap $\Delta_{max}/E_F$ ($\Delta_{max}=\Delta_k$ at $k=0$) as a function of temperature for different values of population imbalance $\alpha$ with $\lambda = 1$ and interlayer separation $l = 0.5r_0$.}
 \label{fig.1.8}
\end{figure}

\begin{figure}[htbp]
\centering
\includegraphics[height=2.1in,width=2.9in]{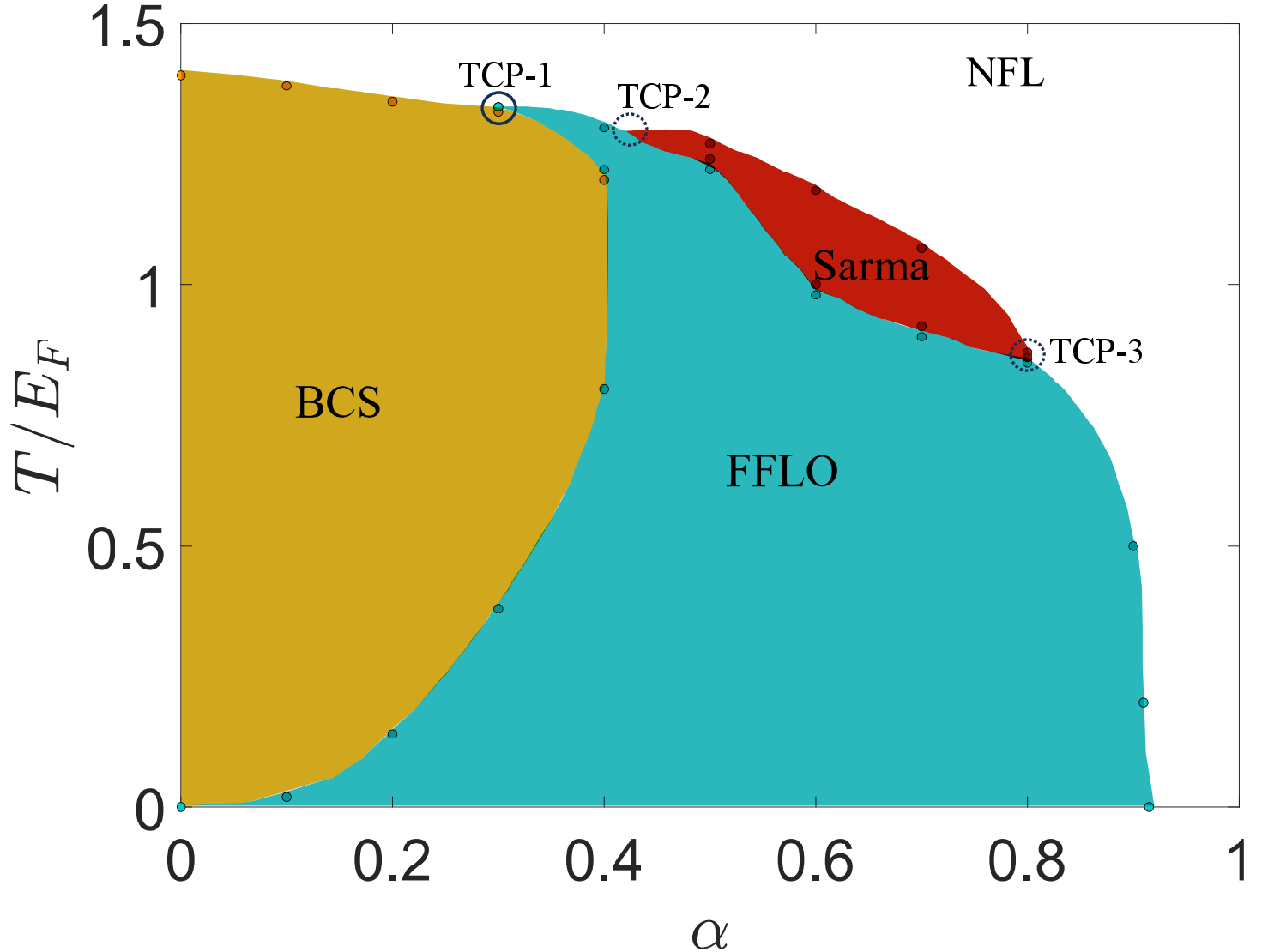}
\caption{Thermodynamic phase diagram of an imbalanced bilayer dipolar system as a function of population imbalance for $l=0.5r_0$ and $\lambda=1$.}
\label{fig.1.9}
\end{figure}
Next we discuss the finite temperature effects on this imbalanced dipolar Fermi gas. In Fig.\ref{fig.1.8} we show the finite temperature dependence of $\Delta_{max}$ (in scale of $E_F$), maximum value of $k$-dependent superfluid gap, for different population imbalance with $l=0.5r_0$ and $\lambda=1$. For vanishing density imbalance, $\alpha = 0$, only BCS phase is present. The parameters are chosen in such a way that the chemical potential is positive and close to zero, which is the unitarity condition for our case. As the imbalance increases, FFLO and Sarma phases start to appear. At $T=0$ and $\lambda=1$, as observed in Ref.\cite{Mazloom_PRB:2017} the chemical potential is close to zero which is similar to the unitarity limit in the contact interaction case. At zero temperature, the BCS (for $\alpha=0$) and FFLO (for $\alpha>0$) are present. However, at finite temperatures just below $T_c$, Sarma phase appears for large density imbalance indicating the signature of two types of tricritical points at finite temperature one corresponds to the coexistance of BCS, FFLO and NFL and the other corresponds to the coexixtance of FFLO, Sarma and NFL phase. All the previous studies on polarized Fermi superfluid with both contact \cite{Gubankova_PRB:2006,Pao_PRB:2006} and dipolar \cite{Mazloom_PRB:2017} interactions have shown the only the stable FFLO phase in the BCS side ($\mu>0$) of a BCS-BEC crossover. The Sarma phase is shown to be only stable deep inside the BEC regime ($\mu<0$). Although this is true at zero temperature, as we show in Fig.\ref{fig.1.8}, near the unitarity limit in the BCS side one may have stable Sarma phase at finite temperatures and large density imbalance.

The thermodynamic phase diagram of these three phases is shown in the plane of temperature $(T/E_F)$ and density imbalance $(\alpha)$ in Fig.\ref{fig.1.9}. We have taken the interlayer separation $l=0.5r_0$ and $\lambda=1$. It is seen that at finite temperatures all three phases are stable in the BCS side with $\mu>0$ which is unlike at $T=0$ where only the BCS and FFLO phases are stable on the BCS side. For lower population imbalance, the BCS phase is stable even for finite values of $\alpha$ at non-zero temperatures. However, as the imbalance increases FFLO phase becomes stable for zero- and finite-temperatures. The Sarma phase is stable only at finite temperature near $T_c$ and large imbalance. As may be seen in the phase diagram, there are three Lifshitz points denoted by TCP-1, TCP-2 and TCP-3 in Fig.\ref{fig.1.9} with corresponding $(T_c, \alpha_c)$ given as TCP-1=$(1.34E_F, 0.3)$, TCP-2=$(1.29E_F, 0.42)$ and TCP-3=$(0.86E_F, 0.8)$. These Lifshitz points are of two types. TCP-1 corresponds to the coexistence of BCS, FFLO and NFL phase while TCP-2 and TCP-3 represent the coexistance of FFLO, Sarma and NFL phase. It may be emphasised here that while the Sarma phase and FFLO phase are unstable for contact interaction, for dipolar interaction they correspond to stable phases for appropriate parameters in the temperature and population imbalance plane. 

\section{Summary and conclusions} \label{sec.4}
In conclusion, we have investigated the thermodynamics of a two-component imbalanced superfluid Fermi gas on the BCS side of a BCS-BEC. Specifically, we have considered imbalance between the population of the two components that leads to nonuniform superfluid phases. We have studied two types of Fermi gases, namely (i) homogeneous Fermi gas with contact interactions and (ii) magnetic dipolar Fermi system. We have used the superfluid mass density criterion (\ref{eq.17}) to determine the stability of the nonuniform phases. For a dipolar Fermi gas, superfluid phase is stable in a 2D or quasi-2D geometry and with dipole moments polarized perpendicular to the plane \cite{Baranov_CR:2012,Zinner_PRA:2012}.

For an imbalanced Fermi gas with contact interaction, a pair of solutions emerges for the superfluid gap for different strengths of interaction and different temperatures. The BCS solution is stable, while the IG or Sarma phase solution is unstable. For a constant population imbalance, increasing interaction strength suppresses the IG phase while enhances the BCS phase. Consequently, the critical temperature for the BCS state increases with interaction strength but decreases for the IG state and ceases to exist as the system approaches a BCS-BEC crossover. 

For a bilayer dipolar Fermi gas, we show that the BCS and FFLO states are stable on the BCS side of a crossover for zero and finite density imbalances, respectively, at zero temperature. A stable Sarma phase at zero temperature with a finite density imbalance exists only on the BEC side of a crossover. However, our study shows that all three phases (BCS, FFLO and Sarma) can be stabilized on the BCS side at finite temperature with proper system parameters. The finite temperature phase diagram indicates that there exist three tricritical points for the bilayer dipolar gas in comparison to the isotropic contact interaction case, which indicates a single tricritical point. On the other hand, as the Sarma phase is stable at a relatively large finite temperature for the dipolar system, it is worth studying such systems experimentally. With recent experimental realization of Feshbach resonance with dipolar and non-dipolar Fermi mixer, one can prepare and study an imbalanced dipolar system by implementing two hyperfine states of a dipolar atomic system.

One can further study a system with more than two layers or change the angle of polarization of the dipolar gas with respect to the plane. With an orientation that is no longer perpendicular to the plane, a stable fermionic superfluidity can still be observed up to a certain critical angle, which depends on the dipolar strength \cite{Baranov_CR:2012}. The Berezinskii-Kostarlitz-Thouless transition \cite{Berezinskii_SPJETP:1972, Kosterlitz_JPC:1973} which has been predicted in bilayer dipolar Fermi gas \cite{Zinner_PRA:2012} can also be studied in imbalanced case with possible nontrivial effects. For the multi-layer case, one might expect the emergence of more interesting topological phases.  

\section{Acknowledgment}
SM is thankful to Indian Association for the Cultivation of Science, Kolkata, India and DST, Govt. of India for research fellowship.
 \bibliography{main2}
 \bibliographystyle{apsrev4-2.bst}
\end{document}